# Generative Adversarial Network Based Synthetic Learning and a Novel Domain Relevant Loss Term for Spine Radiographs

Ethan Schonfeld
Stanford Medicine
Neurosurgical Artificial Intelligence and Machine Learning Laboratory
eschon22@stanford.edu

Anand Veeravagu
Stanford Medicine
Neurosurgical Artificial Intelligence and Machine Learning Laboratory
anand.veeravagu@stanford.edu

## Abstract

*Problem: There is a lack of big data for the training of deep learning models in medicine, characterized by the time cost of data collection and privacy concerns. Generative adversarial networks (GANs) offer both the potential to generate new data, as well as to use this newly generated data, without inclusion of patient's real data, for downstream applications. Approach: A series of GANs were trained and applied for a downstream computer vision spine radiograph abnormality classification task. Separate classifiers were trained with either access or no access to the original imaging. Trained GANs included a conditional StyleGAN2 with adaptive discriminator augmentation (StyleGAN2–ADA), a conditional StyleGAN2 with adaptive discriminator augmentation to generate spine radiographs conditional on lesion type (StyleGAN2–ADA–MultiClass), and using a novel clinical loss term for the generator a StyleGAN2 with adaptive discriminator augmentation conditional on abnormality (SpineGAN). Finally, a differential privacy imposed StyleGAN2–ADA conditional on abnormality was trained and an ablation study was performed on its differential privacy impositions. Key Results: We accomplish GAN generation of synthetic spine radiographs without meaningful input for the first time from a literature review. We further demonstrate the success of synthetic learning for the spine domain with a downstream clinical classification task (AUC of 0.830 using synthetic data compared to AUC of*

*0.886 using the real data). Importantly, the introduction of a new clinical loss term for the generator was found to increase generation recall as well as accelerate model training. Lastly, we demonstrate that, in a limited size medical dataset, differential privacy impositions severely impede GAN training, finding that this is specifically due to the requirement for gradient perturbation with noise.*

## 1. Introduction

Artificial intelligence, specifically machine learning, has transformed countless fields through task automation or by decision making support. However, while medicine has benefited greatly from machine learning, an example of such would be the release of the CheXpert dataset leading to successful development of chest radiograph classifiers [7] [18], it has not experienced the same machine learning driven advancements as other fields have, largely due to a lack of data. Contributing factors to this lack of data are i. the high time cost of data collection and ii. privacy concerns. These privacy concerns have hampered large dataset creation and data sharing across centers, which is necessary to prevent distributional shift. This paper develops the emerging concept of synthetic learning, the training of generative adversarial networks (GANs) to generate data of sufficient quality for downstream model development without sharing the true medical data for privacy protection. In this paper we apply synthetic learning to spine imaging, demonstrate that synthetic learning is possible for downstream clinical and diagnostic applications, introduce a new clinical loss term for the generator that was found to increase generation recall as well as accelerate model training, and finally demonstrate the challenge of using differential privacy for GAN training on a limited sized medical dataset.

## 2. Background and Related Work

While Imagenet, used for training state of the art computer vision models, contains more than 14 million images, in the spine imaging domain such large scale data sources are non-existent [4]. However, large scale data volume is required for successful machine learning [5]. In 2020, the largest spine imaging dataset that had been released was composed of only 797 images, and the second largest of only 60 images [9] [10] [13]. In 2021, VinDR–SpineXR released an imaging dataset of 10,469 spine radiographs labeled for multiple abnormalities and lesions with bounding boxes [14]. VinDR–SpineXR also released an accompanying deep learning framework that achieved an AUC of 0.886 for classifying these spine radiographs as normal or abnormal, which is state of the art performance [13]. This novel release provides both the training imaging data, and a baseline for model performance, for the validation of synthetic learning in the spine radiograph domain.

Synthetic learning aims to develop privacy preserving generating models that can learn the real image distribution and generate corpuses of images capable of training downstream models. In this paper, we will first focus on demonstrating synthetic learning feasibility without privacy preservation as this has yet to be achieved in spine imaging or achieved for downstream clinical classification tasks. Synthetic learning has been demonstrated on segmentation models for brain and nuclei imaging [3]; however, from a literature review, has not been demonstrated for medical clinical classification tasks. Relevant work includes GAN based data augmentation leading to improved liver lesion classification; however, this work included the training data in the classifier training [6]. A similar approach used a DC–GAN to generate multi conditional cell types and showed that inclusion of this synthetic data improved classification model training accuracy, but did not demonstrate model success training on only these synthetic images [12]. In this cell type classifier, when real data inclusion is reduced by 80%, the GAN supplemented model, using more synthetic images, has decreased performance [12]. Thus, we intend to study if decreasing real data inclusion by 100%, using only synthetic data, for a more difficult clinical abnormality classification task, significantly harms classifier performance.

Privacy preserving generative methods include: federated learning, split learning, and asynchronized discrimination [3]. Federated learning achieves privacy by using local nodes to learn a global model and privately communicates this information by adding random noise to gradients / parameters [2] [3]. Split learning does so by using a data block to cut layer gradients and isolating central learning from direct data contact [17]. Finally, asynchronized discrimination does so by using fake data and an auxillary variable and discriminator loss [3]. The need for this privacy comes from findings that deep learning models trained on image data could be used to reconstruct that sensitive training data [11]. Such adversarial attacks are often carried out by performing inference on model gradients [11]. Thus, one main approach towards protection against such attacks, known as differential privacy, uses a non–convex objective to train deep learning networks under a modest privacy budget [1]. Differential privacy imposes: clipping the L2 norm of every gradient, using microbatch (small sized minibatches) achieving best privacy under individual batch sizes, and adding noise to batch gradients before all parameter update steps [1].

## 3. Dataset

VinDR-SpineXR dataset has 10,466 spine radiographs that are annotated for 13 different classifications with respective bounding boxes [14]. These radiographs, coming from the Hanoi Medical University Hospital, were labeled for the 13 lesions by a committee of three experienced radiologists. Example classifications are enthesophytes, vertebral collapse, or spondylolysthesis. Each of the 5,000 studies included in the dataset was labeled by one of the three radiologists. For each imaging study, information on the presence of an abnormality, the classification of the abnormality, bounding boxes of each abnormality, and basic demographic information of the patient is provided. This data was downloaded from PhysioNet after signing the access policy. The dataset is split into a training set that contains 8,389 images and a test set that contains 2,077 images. Each image is in DICOM format with some identifying information removed. The pixel arrays of the DICOM images are grayscale and of varying dimensions and resolutions. Because DICOMS were identified, many are missing information, some even missing pixel arrays, and thus require serious preprocessing. Each DICOM was preprocessed to extract its pixel array and its abnormality status using the image ID. To further preprocess each pixel array, the smallest dimension was taken of the image, and a square, centered at the image's center was cropped for final use from the pixel array. Any image with the smallest dimension less than 128 pixels was not used for model training. Finally, all square cropped images were downsampled using a nearest neighbors approach to 256 by 256 resolution. After preprocessing, there were 1470 abnormal radiographs and 2303 normal radiographs to be used for all later model training (Figure 4a).

## 4. Methods

### 4.1. Preprocessing and Training

All training was done on a google cloud virtual machine using 1 x NVIDIA Tesla A100. Data augmentation for the classification models included a random horizontal flip with 0.3 probability, a random rotation from -5 to 5 degrees, and a random resized crop (scale=0.8, 1.0) and (1.0, 1.0) ratio. Preprocessing for all classification models included a resize to 224 square pixels, and normalization to (mean=[0.485,
0.456, 0.406], std=[0.229, 0.224, 0.225]).

### 4.2. Classifier Models

For binary abnormality classifier (SpineClassifier) experiments, the DenseNet121 model architecture was used with a linear layer, dropout (0.1), and sigmoid activation added to the end of the model. Weights were pretrained on Imagenet. Binary cross entropy loss with class weights was used as the loss function. Training continued for all classifiers for 30 epochs. Classifiers were trained augmented with data as either i. Standard augmentations, ii. 500 (250 normal, 250 abnormal) StyleGAN–ADA generated images added to the training set, iii. 500 (250 normal, 250 abnormal) SpineGAN generated images added to the training set, iv. 1600 (800 normal, 800 abnormal) StyleGAN–ADA generated images added to the training set, v. 1600 (800 normal, 800 abnormal) SpineGAN generated images added to the training set, vi. 10,000 (5,000 normal, 5,000 abnormal) StyleGAN–ADA generated images added to the training set, or vii. 10,000 (5,000 normal, 5,000 abnormal) SpineGAN generated images added to the training set, in order to study the effect of GAN loss function, generative augmentation amount effect on classifier performance as compared to traditional standard augmentation methods, and synthetic training.

### 4.3. GAN Models

All GAN models used the architecture of StyleGAN2–ADA as adaptive discriminator augmentation was found to improve generative quality for small sized training datasets [8]. StyleGAN2–ADA was made to be a conditional GAN for 2 classes: normal spine radiographs (2303 true images) and abnormal spine radiographs (1470 true images). StyleGAN–ADA–MultiConditional was trained to be conditional with eight classes of radiographs: No finding, Disc space narrowing, Foraminal stenosis, Osteophytes, Spondylolysthesis, Surgical implant, Vertebral collapse, Other lesions. Any true spine radiographs with more than one of the types of above were not included in GAN training; there were 494 of these multiple abnormality type radiographs. Data augmentation for all GAN models included adaptive discriminator augmentation (parameter 0.6) (blit, geom, color, filter, noise, cutout) enabled. Both StyleGAN2–ADA and SpineGAN were trained for 4,200,000 images shown to the discriminator; however, due to compute costs StyleGAN–ADA–MultiConditional was trained with only 1,200,000 images shown to the discriminator. StyleGAN2–ADA is known to take longer than 3 days to converge; however, as proof of concept for privacy generation and the domain loss term, this paper did not train until convergence.

### 4.4. Differential Privacy

Finally, differential privacy was implemented using the StyleGAN2–ADA backbone architecture. This implementation was done in such a way that an ablation study could be performed. Minibatch size was set to 1 (microbatch) in order to later clip gradients on an instance basis. Gradient clipping (L2) of norm size 1.5 was enforced for all gradients during training. Because noise multiplication requires advanced implementation and a fine tuning of this hyperparameter for decent results, the experiment shown in Figure 1 was performed to estimate a noise level to optimize privacy budget while allowing training for 4,200,000 to mirror the training time of SpineGAN. Figure 1 demonstrates the relationship between noise level and the epsilon value, which is an upper bound of the privacy budget: a measurement of the strength of the privacy guarantee. The delta level was chosen as the inverse of the number of training examples [15], 0.000265. The delta level is a bound on the probability of the privacy guarantee not holding. The tensorflow–privacy library was used to calculate the resulting epsilon value which was then used to determine the optimal noise level. Noise multiplier value is estimated according to the ratio of the noise standard deviation to the gradient clipping [16]. Seeking to choose the lowest possible noise level while maintaining a strong upper bound for privacy budget, noise levels corresponding to using gaussian noise standard deviations of (0.03, 0.06, 0.1, 0.2, 0.3, ..., 1.5) were considered. From Figure 1, the value of 0.2 noise standard deviation was chosen to optimize epsilon while minimizing the added noise to gradients to encourage privacy at the minimum damage to training. Noise multiplication of gradients was implemented with the addition of a random value from the standard normal deviation scaled by a 0.2 hyperparameter for each gradient value. The ablation study trained four Differential Privacy GANs: one as above, one with a 0.02 noise hyperparameter to isolate the noise on gradient effect, one with a batch size of 4 to isolate the individual batch size effect, and one with no gradient norm clipping to isolate this effect on spine radiograph generation.

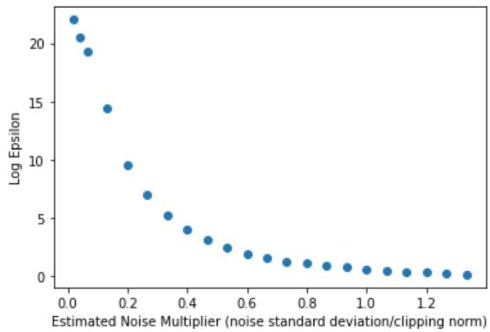

Figure 1. Determination of noise standard deviation for Differential Privacy GAN based on optimization of minimizing epsilon with minimization of necessary added noise standard deviation. Standard deviation of 0.2, corresponding to a 0.1333 estimated noise multiplier was selected (fourth point from left).

4.5. SpineGAN Novel Loss

SpineGAN differs from StyleGAN2–ADA in its loss function. SpineGAN loss function uses the negative log sigmoid of the generated logits for the main generator loss function as is used in StyleGAN2–ADA; however, it also adds a term to this loss function. First, SpineClassifier was trained for 19 epochs with the above standard augmentation. After every generation in the SpineGAN main loop, SpineClassifier is applied to the generated images and Binary Cross Entropy loss term using the true labels as the conditional inputs is added to the main generator loss term for SpineGAN. This additional BCE term was weighted with an empirically determined 0.01 factor. This additional loss term was developed in order to provide knowledge of abnormality during GAN training. While in theory, a conditional GAN should learn both how to generate radiographs as well as condition them on abnormality, because spine radiographs differ in such few features between these conditions, it was hypothesized that including this extra loss term could accelerate the GAN's training. With S as a pretrained network to classify a spine radiograph as normal or abnormal (in this paper with test AUC of 0.856) and c as a vector of the class labels for an m sized minibatch,

$$\text{Let } \hat{x} = G(\hat{c})$$

$$G_l = -\log\sigma(D(\hat{x},\hat{c})) + \gamma \sum_{i=1}^{m} c_i \log S(\hat{x}_i) - (1-c_i)\log(1-S(\hat{x}_i))$$

## 4.6. Generative Metrics

Checkpoints to the GAN models were saved after every 200,000 images were shown to the discriminator. Frechet´ Inception Distance (FID), Precision, and Recall metrics were computed using the training data as the reference for these metrics.

## 5. Experiments 6. Analysis

| Abnormality Classification | Test AUC |
| --- | --- |
| Nyugen et al [13] (State of the Art) | 0.886 |
| True Images | 0.856 |
| True Images + StyleGAN2–ADA (500) | 0.843 |
| True Images + SpineGAN (500) | 0.850 |
| True Images + StyleGAN2–ADA (1600) | 0.847 |
| True Images + SpineGAN (1600) | 0.856 |
| True Images + StyleGAN2 (10000) | 0.839 |
| True Images + SpineGAN (10000) | 0.850 |
| StyleGAN2–ADA (10000) | 0.814 |
| SpineGAN (10000) | 0.830 |

Table 1. Performance on abnormality classification task using standard augmentation, domain augmentation, and synthetic training

| Generation Quality | Frechet Inception Distance´ | Precision | Recall |
| --- | --- | --- | --- |
| StyleGAN2–ADA | 32.7 | 0.460 | 0.0159 |
| SpineGAN | 33.9 | 0.418 | 0.0194 |

Table 2. Final GAN performance metrics for the fully trained networks (4,200,000 images shown to the discriminator).

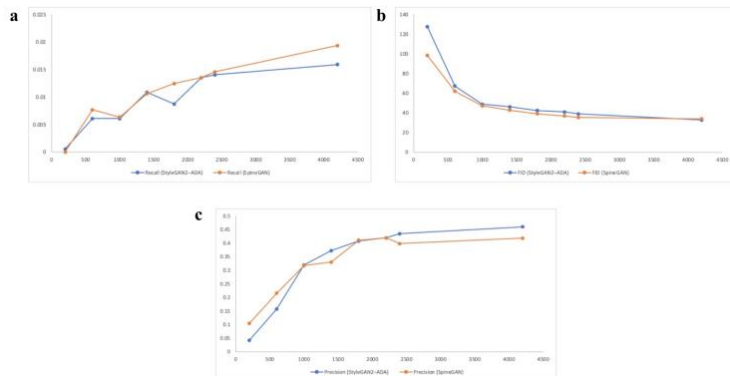

Figure 2. GAN Performance by increasing training checkpoints (in thousands of images shown to the discriminator). SpineGAN differs from StyleGAN2–ADA by an additional domain loss term using a trained abnormality classifier. This domain loss term appears to accelerate training while resulting in similar (a. Frechet´ Inception Distance (FID), b. Precision) or greater (c. Recall) outcomes. As an experiment, SpineGAN was trained for 200,000 images with the loss term weighted 5 times heavier (0.05 vs. 0.01), and FID, Precision, and Recall metrics were respectively: 147.08, 0.0292, and 0.0007951 which demonstrates that while the addition of the domain loss term accelerates training, weighing this term too heavily can dominate feedback to the generator and diminish discriminator feedback which then likely slows training as evidenced by these 200,000 timestep metrics. Blue is SpineGAN and Orange is StyleGAN2–ADA

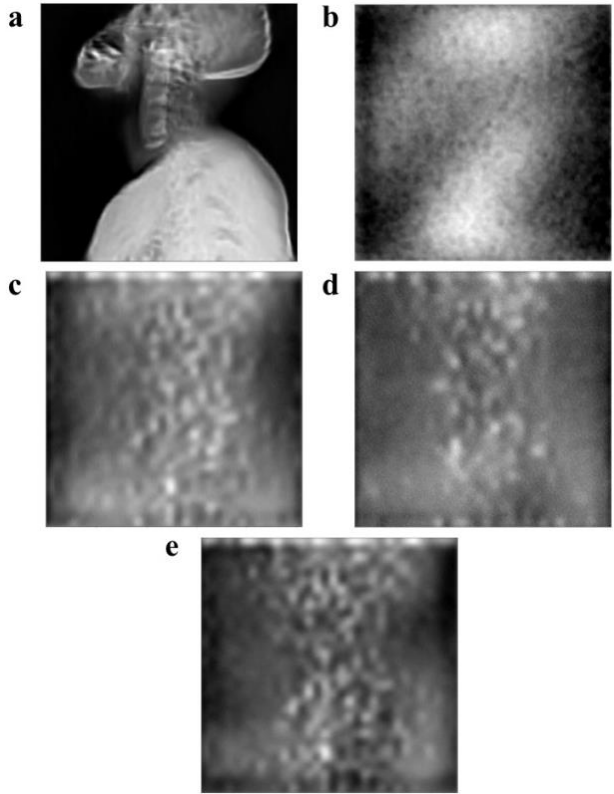

Figure 3. Ablation study of differential privacy generated outputs at training time of 200,000 images shown to the discriminator. a. SpineGAN, b. Differential privacy with 0.2 noise factor for gaussian noise addition for gradients, c. Differential privacy with 0.02 noise factor, d. Differential privacy with minibatches of 4 instead of 1, e. Differential privacy without L2 norm (1.5) clipping gradients

Successful synthetic training is demonstrated in Table 1 on the abnormality classification task. State of the art for this task is 0.886, a convolutional network classifier trained on real data achieved 0.856, and a convolutional network classifier trained only on synthetic data achieved 0.830. Table 1 further demonstrates that for the standard StyleGAN2–ADA synthetic data augmentation, performance increases as the synthetic amount of images is raised from 500 to 1600, but ultimately decreases when including 10000 images. This could be due to a low recall of the generative model (Table 3). Using SpineGAN, with the addition of the novel clinical abnormality classifier loss term for the generator, augmentation performance increases as the synthetic amount of images increases, despite with 10000 images, however, with a lesser effect. This is further evidenced as the recall of SpineGAN after full training is greater than that of the baseline loss function StyleGAN2–ADA (Table 3). Finally, the classifier trained on SpineGAN synthetic images outperforms the classifier trained on StyleGAN2–ADA synthetic images (Table 1). GAN generated outputs after full training are visualized in Figure 4.

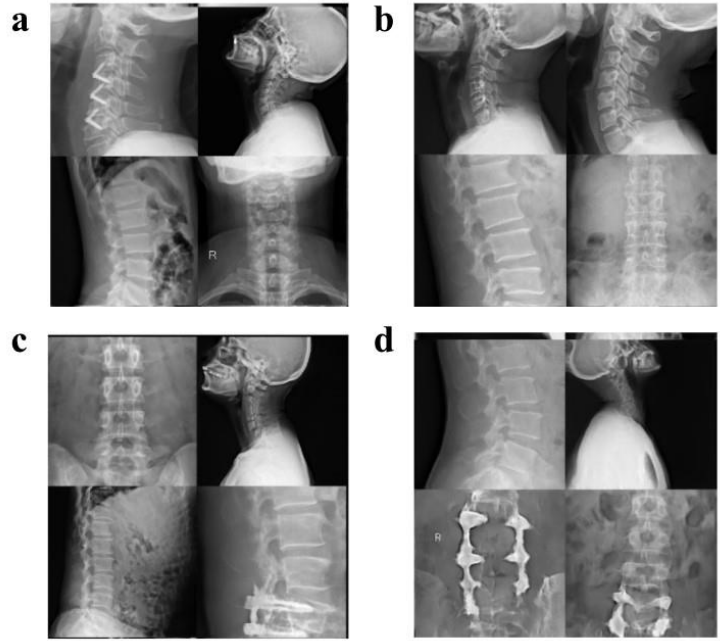

Figure 4. a. Real spine radiographs used during GAN training, and Visualized generated outputs from b. StyleGAN2–ADA, c. SpineGAN, d. StyleGAN2–ADA–MultiConditional, without truncation.

Comparing StyleGAN2–ADA with SpineGAN offers evidence that the additional clinical term not only accelerates the GAN training (Figure 2), but also results in modeling a different distribution. While the Frechet Iń- ception Score (FID) is greater for the fully trained StyleGAN2–ADA, SpineGAN has a similar FID score while having a greater recall score (Table 3). This difference in recall scores can be seen to increase in differential as the two GANs are trained for longer, suggesting that if recall and image diversity is important for a medical GAN, the inclusion of a domain relevant classifier as a loss term can accomplish this.

Empirically, the multi conditional GAN was verified as being able to generate consistently the desired classes of lesions. The bottom two images of Figure 4d demonstrate success of StyleGAN2–ADA–MultiConditional generations using the class vector of surgical implant condition.

The ablation study in Figure 3 demonstrates that the addition of gaussian noise to gradients during training results in poor generative quality in the case of using a limited size training set of spine radiographs. Figure 3 shows that despite allowing increased batch size (d) and no restriction on gradient norm (e) that as long as there is addition of Gaussian noise to gradients (b–e) that generative quality is poor. This is further confirmed by increasing the scaling factor of this Gaussian noise (b) which results in increased damage to the generative quality.

## 7. Conclusion

In this paper we demonstrate a proof of concept for synthetic learning on spine radiographs for downstream clinical classification tasks. We further introduce a novel loss function and show that it accelerates training and increases recall. Lastly, we provide evidence that among the three requirements for differential privacy during training, the noise addition to gradients is what significantly hinders the generative model training. Future work can develop differential privacy to relax the requirement of noise addition to gradients and instead seek to prevent gradient inference by other methods.

The loss function we include is scalable to other medical domain generative efforts. The requirement is a high performing classifier on the true images for some relevant classification task, not confined to solely the classification task that the synthetic learning is intended for. Future work should systematically investigate the selection of appropriate classification tasks that best accelerate model training or best increase recall, as well as the domain dependency of these tasks.

## 8. Declarations

Disclosure of funding, financial support, industry affiliation: AV: Consultant – Medtronic, Stryker, Nuvasive, Surgical Theater, Osteocentric; Other authors: None

Conflicts of Interest: None